\documentclass{PoS}
\usepackage{graphicx}

\title{Exclusive $J/\Psi$ photo- and electroproduction in a dual model}

\ShortTitle{$J/\Psi$ photo- and electroproduction in a dual model}

\author{\speaker{A.~Prokudin}\\
        Jefferson Lab,12000 Jefferson Avenue, Newport News, VA 23606\\
        E-mail: \email{prokudin@jlab.org}}

\author{R. Fiore\\
        Dipartimento di Fisica, Universit\`{a}
della Calabria and \\ Instituto Nazionale di Fisica
Nucleare, Gruppo collegato di Cosenza\\
I-87036 Arcavacata di Rende, Cosenza, Italy}

\author{L.L. Jenkovszky \\
Bogolyubov Institute for Theoretical Physics (BITP),
Ukrainian National Academy of Sciences \\14-b, Metrolohichna str.,
Kiev, 03143, Ukraine}

\author{V.K.~Magas \\
Departament d'Estructura i Constituents de la
Mat\'eria,\\ Universitat de Barcelona, Diagonal 647,\\
08028 Barcelona, Spain
}

\author{S.~Melis \\
Di.S.T.A., Universit\`a del Piemonte Orientale
``A. Avogadro'', 15100
Alessandria, Italy, \\
INFN, Gruppo Collegato di Alessandria, 15100
Alessandria, Italy 
}

\abstract{Exclusive $J/\Psi$ electroproduction is studied in the framework
of the analytic $S-$matrix theory. The differential and integrated
elastic cross sections are calculated using the Modified Dual
Amplitude with Mandelstam Analyticity (M-DAMA) model. The model is
applied to the description of the available experimantal data and
proves to be valid in a wide region of the kinematical variables
$s$, $t$ and $Q^2$. Our amplitude can be used also as a universal
background  parametrization
for the extraction of tiny resonance signals}

\FullConference{XVIII International Workshop on Deep-Inelastic Scattering and Related Subjects, DIS 2010\\
		April 19-23, 2010\\
		Firenze, Italy}

\begin{document}

\newcommand{\dlt}{\bigtriangleup}
\newcommand{\beq}{\begin{equation}}
\newcommand{\eeq}[1]{\label{#1} \end{equation}}
\newcommand{\insertplot}[1]{\centerline{\psfig{figure={#1},width=14.5cm}}}

\parskip=0.3cm

\section{Introduction} \label{s1}

\begin{figure}[h]
\centering \includegraphics[width=.45\textwidth,bb= 72 471 470 670]{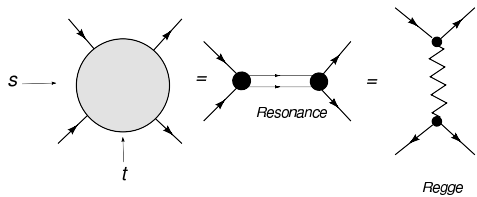}
\caption{Dual description of hadron scattering amplitude.}
\label{fig_1}
\end{figure}

Duality in strong interaction physics is the relationship between the description of hadronic scattering amplitudes in terms of $s$-channel resonances at low energies, and $t$-channel Regge
poles at high energies, see Fig.~\ref{fig_1}, for review see Ref.~\cite{Melnitchouk:2005zr}.
The Veneziano model ~\cite{Veneziano} embodied duality and correct Regge asymptotic behaviour. The dual model with Mandelstam analyticity (DAMA) \cite{DAMA} appeared as a
generalization of Veneziano dual model. Contrary to narrow-resonance dual models, DAMA requires  non-linear, complex
trajectories. The dual properties of DAMA were studied
in Ref. \cite{Jenk}. High-energy exclusive $J/\Psi$ electroproduction was intensively
studied in recent years, for a review see Ref.~\cite{review}. One of the most interesting properties is the suppression of
the so-called secondary reggeon exchanges resulting from the Okubo-Zweig-Iizuka rule \cite{OZI}.  $J/\Psi$ photoproduction is an ideal tool to study the
Pomeron exchange. The leading Pomeron exchange can be modelled using DAMA \cite{DAMA} and Modified DAMA \cite{MDAMA} amplitudes. The main source of the data at
high energies is the HERA collider at DESY~\cite{breitweg99,adloff99,chekanov02,chekanov04,aktas06}, while
at low energies experimental studies at JLab are promising, see
for instance the recent paper of Ref.~\cite{JLab}. In Refs.~\cite{our_paper,Fiore:2009xk} a model based on DAMA and MDAMA
 was proposed to describe $J/\Psi$ photo- and electro-production; it
resulted in a very good description of the data in the region of
$s$, $Q^2$ from the threshold to high energy, where pure Pomeron exchange
dominates.

\section{The model} \label{s2}

The DAMA amplitude \cite{DAMA} is given
by:
\beq D(s,t)=c \int_0^1 {dz \biggl({z \over g}
\biggr)^{-\alpha(s')-1} \biggl({1-z \over
g}\biggr)^{-\alpha_t(t'')}}\,,
\eeq{dama_eq}
where $\alpha(s)$ and
$\alpha(t)$ are Regge trajectories in the $s$ and $t$ channel
correspondingly,  $g$ and $c$ are
parameters, with $g>1$, $c>0$, and the notations $x'=x(1-z), \ \ x''=xz$ $(x=s,\, t, \, u)$ are used.

To extend the model off-mass shell we need to construct the $Q^2$-dependent dual amplitude. To this aim we use
the so called Modified DAMA (M-DAMA) formalism developed in Ref.~\cite{MDAMA}.
The scattering amplitude is given by
\beq
D(s,t,Q^2)=c \int_0^1 {dz \biggl({z \over g}
\biggr)^{-\alpha(s')-\beta({Q^2}'')-1} \biggl({1-z \over
g}\biggr)^{-\alpha_t(t'')-\beta({Q^2}')}}\,,
\eeq{mdama_eq} where
 $\beta(Q^2)$ is the following monotonically
decreasing dimensionless function of $Q^2$;
$x'=x(1-z), \ \ x''=xz$, where $x \equiv s, \, Q^2, \, t$: 
\beq \beta(Q^2)=-{\alpha_t(0)\over \ln g}\ln
\left(\frac{Q^2+Q_0^2}{Q_0^2}\right)\,. \eeq{beta}
Clearly at $Q^2 = 0$ we have
$\beta(0) = 0$ so that we reproduce the on-mass shell amplitude
of Eq.~(\ref{dama_eq}).

In Refs. \cite{our_paper,Fiore:2009xk} these amplitudes were used to describe $J/\Psi$ photo and electroproduction. Vector meson dominance \cite{VMD} was employed to relate
photon $\gamma^* p \rightarrow J/\Psi p$ and hadron amplitudes $J/\Psi p \rightarrow J/\Psi p$.

 \beq
 A(\gamma^*\, p\rightarrow J/\Psi\, p)=\frac{e}{f_{J/\Psi}} A(J/\Psi\, p\rightarrow J/\Psi\, p),
 \eeq{amplitude}
where $f_{J/\Psi}$ is the decay constant of ${J/\Psi}$. ${e}/{f_{J/\Psi}}$ is included into the parameter of the model $c$.

The amplitude reads
\beq
A(s,t,u,Q^2)=(s-u)(D(s,t,Q^2)-D(u,t,Q^2)).
\eeq{}

Trajectories are chosen in the following way:
\beq
\alpha(s)=\alpha(0)+\alpha_1(\sqrt{s_0}-\sqrt{s_0-s})\, .
\eeq{trajectory}
\beq
\alpha(t)=\alpha^P(t)=\alpha^P(0)+\alpha^P_1(\sqrt{t_1}-\sqrt{t_1-t})+
2\alpha^P_2(t_2-\sqrt{(t_2-t)t_2}). \eeq{Pomer}

The transverse differential cross section is given by
\beq
{d\sigma_T\over{dt}}(s,t,Q^2)={1\over16\pi\lambda (s,
m_{J/\Psi}^2, m_P^2)}\vert A(s,t,u,Q^2)\vert ^2\,,
\eeq{dsigma}

The total elastic cross section is given
by the sum of longitudinal and transverse components: \beq
\sigma_{el}(s,Q^2)=(1+R(Q^2))\sigma_T(s,Q^2)\,, \eeq{sigma_el}
where $R=\sigma_L/\sigma_T$ parametrized as in Ref.~\cite{martynov}:

\beq
R(Q^2)=\left(\frac{a\, m_{J/\Psi}^2 + Q^2}{a\,
m_{J/\Psi}^2}\right)^{n}-1\,, 
\eeq{R}

\begin{table}[t]
\begin{center}
\begin{tabular}{|rl|rl|}
\hline
\multicolumn{4}{|c|}{{\bf ON-MASS-SHELL}  Ref.~\cite{our_paper}}\\
\hline
$\alpha^E(0)$ = & $-1.83$ & $\alpha^E_1(0)$ = & $0.01$ (GeV$^{-1}$) \\
$\alpha^P(0)$ = & $1.2313$ & $\alpha^P_1(0)$ = & $0.13498$ (GeV$^{-1}$)\\
$\alpha^P_2(0)$ = & $0.04$ (GeV$^{-2}$) & $t_2$ = & $36$ (GeV$^2$) \\
$g$ = & $13629$ & $c$ = &  $0.0025$ (GeV$^{-2}$)  \\
\hline
\multicolumn{4}{|c|} {$\chi^2/{d.o.f.}$ =  $0.83$}\\
\hline
\multicolumn{4}{|c|} {{\bf OFF-MASS-SHELL} Ref.~\cite{Fiore:2009xk}}\\
\hline
\multicolumn{4}{|c|} {$Q_0^2$ = $3.464^2$ (GeV$^2$),  $\quad a$ =  $2.164$,  $\quad n$ =  $2.131$ }\\
\hline
\multicolumn{4}{|c|} {$\chi^2/{d.o.f.}$ =  $1.2$}\\
\hline
\end{tabular}
\end{center}
\vspace{-0.5cm}
\caption{Fitted values of the adjustable parameters
\label{fitpar}}
\end{table}

\begin{figure}{}
\includegraphics[width=.45\textwidth,bb= 10 140 540 660]{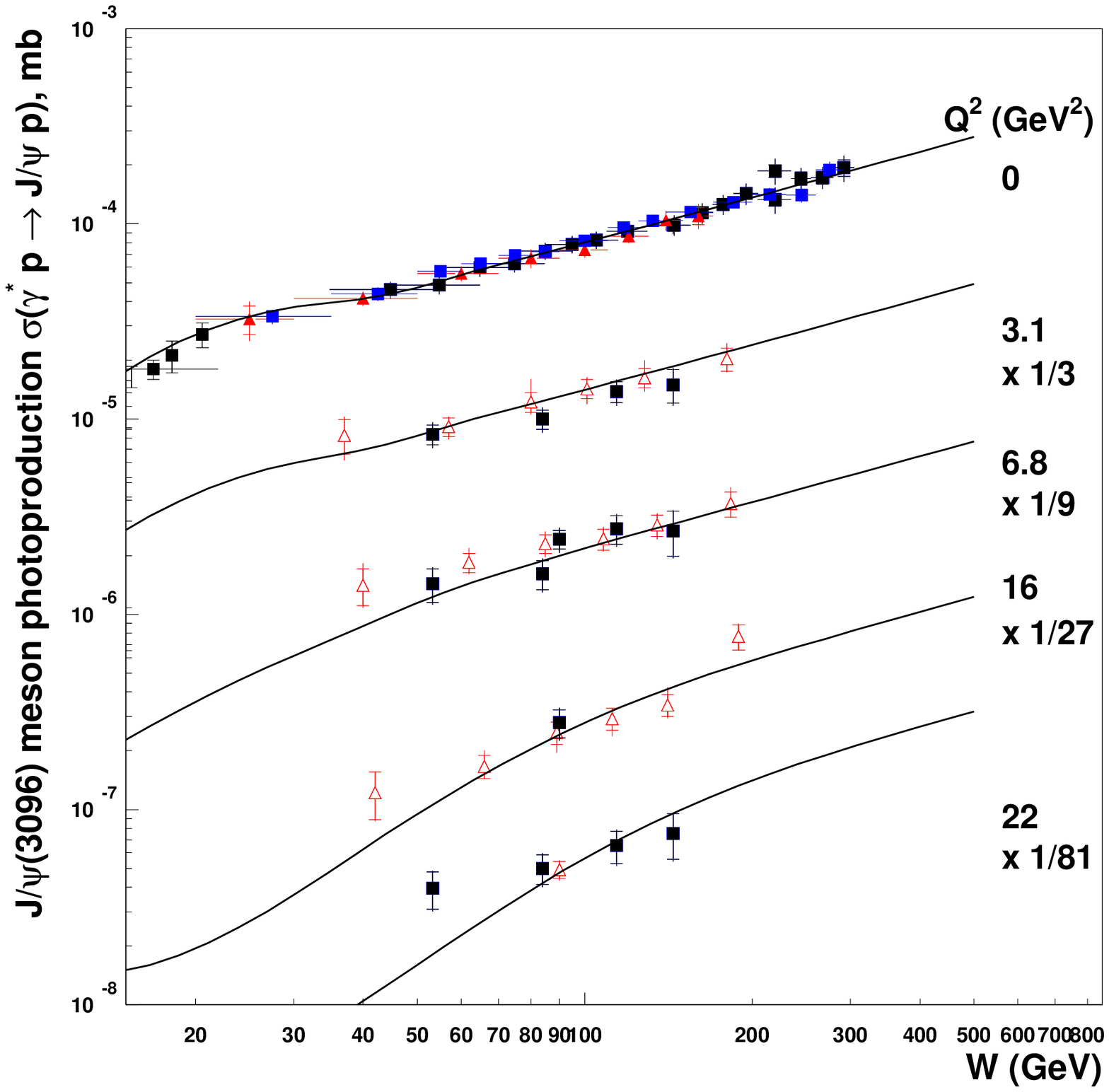}
\includegraphics[width=.45\textwidth,bb= 10 140 540 660]{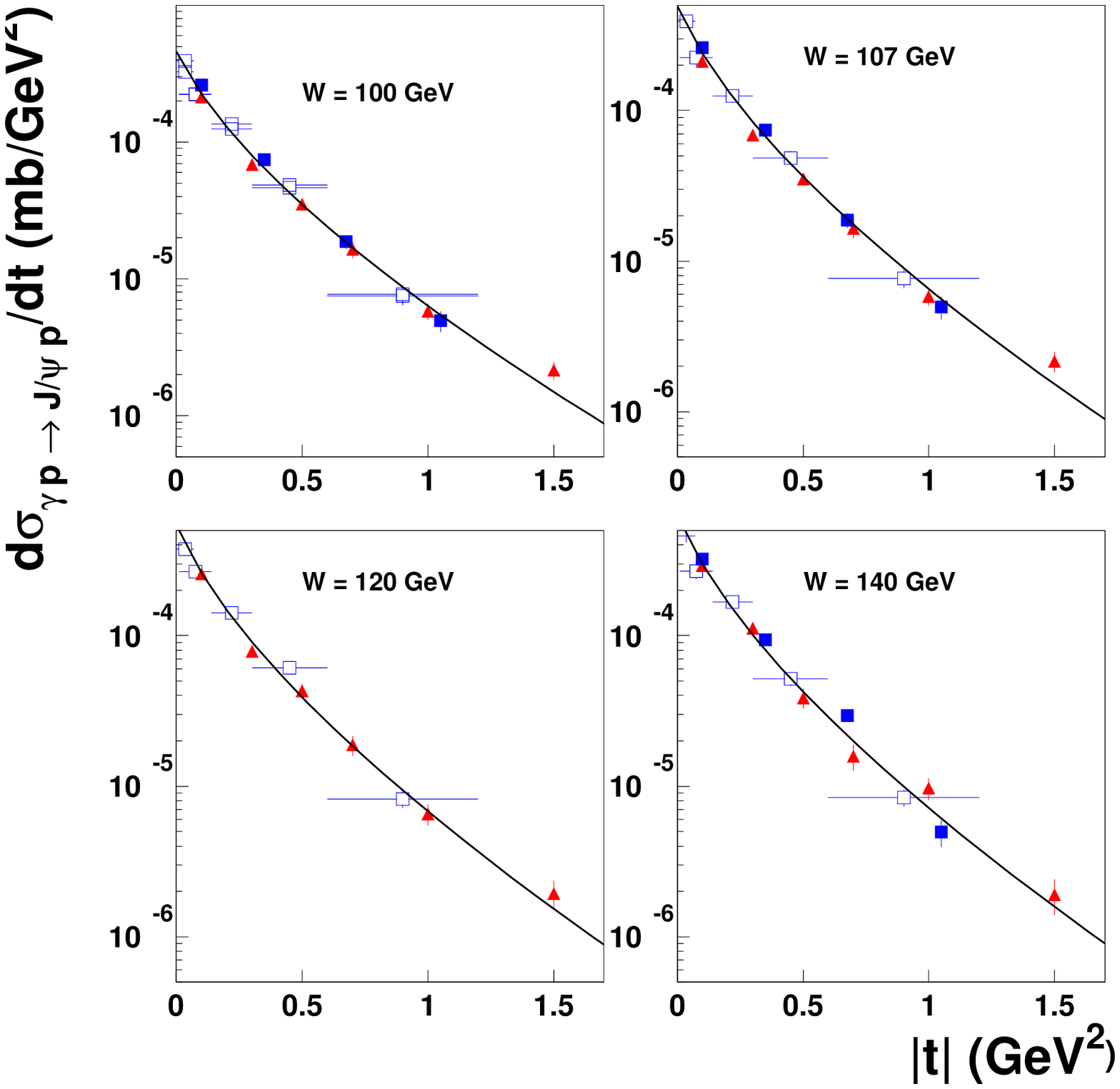}
\caption{$J/\Psi$ elastic and differential cross sections.}
\label{fig_2}
\end{figure}

\begin{figure}{}
\includegraphics[width=.45\textwidth,bb= 10 140 540 660]{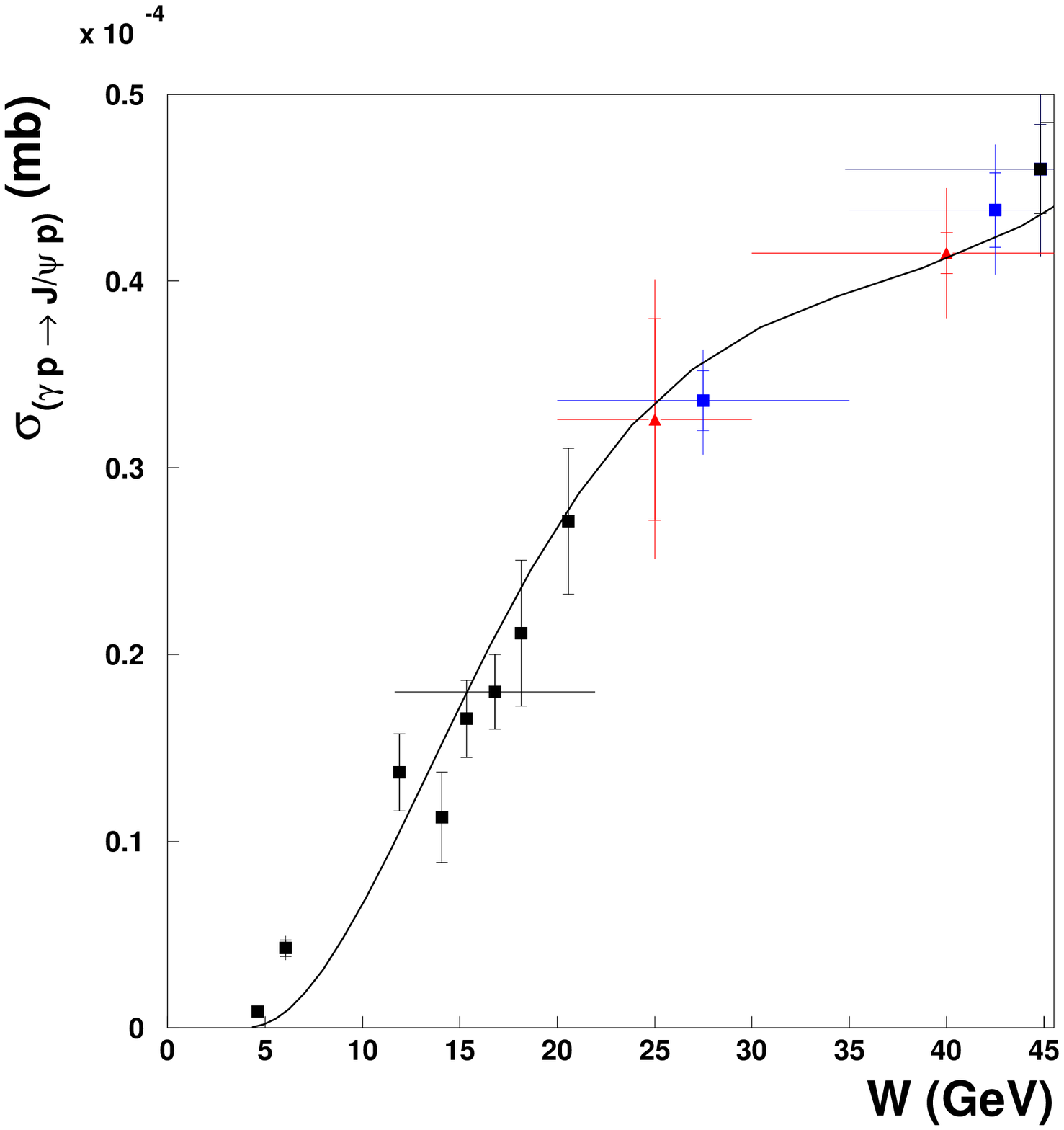}
\includegraphics[width=.45\textwidth,bb= 10 140 540 660]{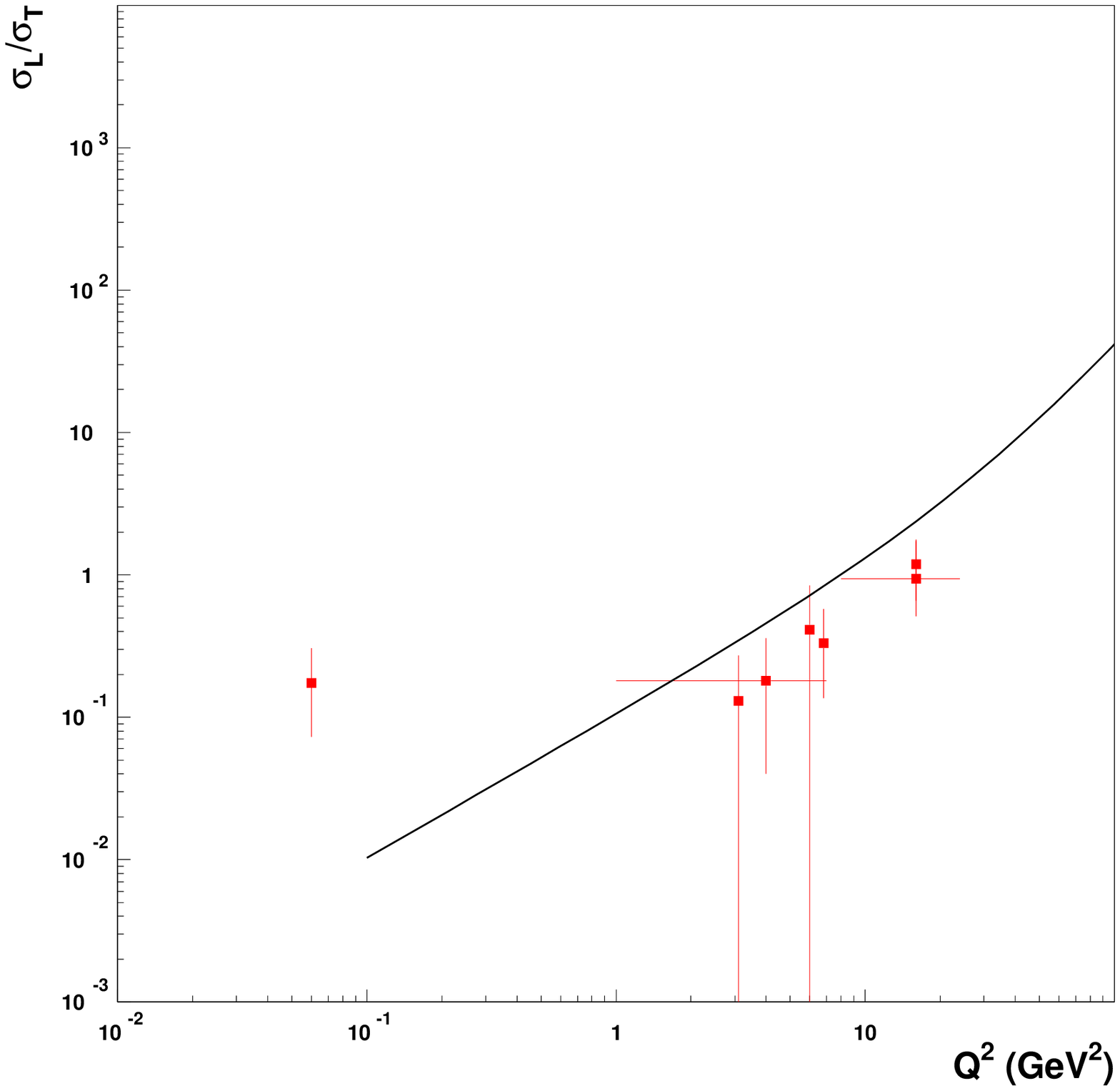}
\caption{$J/\Psi$ elastic cross section at low $W$ and $R=\sigma_L/\sigma_T$ as a function of $Q^2$.}
\label{fig_3}
\end{figure}

The results are presented in Table~\ref{fitpar} and Figs~\ref{fig_2},\ref{fig_3} (see \cite{our_paper,Fiore:2009xk} 
for more results). We can conclude that the agreement is rather good and our 
model has correctly captured the $s$,$t$ and $Q^2$ behaviour of $J/\Psi$ meson 
production.

\section{Conclusions}
A model \cite{our_paper,Fiore:2009xk} that describes $J/\Psi$ photo and electroproduction based on DAMA and M-DAMA~\cite{DAMA, MDAMA} is constructed. 
Good fits of the data are achieved in all available regions of
the Mandelstam variables and photon virtualities. The model can be
used to predict cross sections for $J/\Psi$ photo- and
electroproduction for any value of the Mandelstamian variable, and
in a wide range of photon virtualities $Q^2$.

This model can be applied to other vector meson production. Careful study of the influence of the virtuality of the photon and application of VDM is needed to improve the quality of the model.

\end{document}